\documentclass[runningheads]{svmult}

\usepackage{makeidx}   
\usepackage{graphicx}  
\usepackage{subeqnar}  
\usepackage{multicol}  
\usepackage{physprbb}  
\makeindex             

\newcommand{\greeksym}[1]{{\usefont{U}{psy}{m}{n}#1}}
\newcommand{\umu}{\mbox{\greeksym{m}}}

\begin{document}
\title*{
Dynamics of Magnetization Reversal in Models of Magnetic Nanoparticles
and Ultrathin Films
}
\toctitle{
Dynamics of Magnetization Reversal in Models of
\protect\newline Magnetic Nanoparticles and Ultrathin Films
}
%
%
\titlerunning{Magnetization Reversal in Model Nanoparticles and Films}
%
\author{
Per Arne Rikvold\inst{1,2}
\and Gregory Brown\inst{2,3}
\and Steven~J.\ Mitchell\inst{1,2}
\and M.~A.~Novotny\inst{4}
}
\authorrunning{P.A.\ Rikvold et al.}
%
%
\institute{
Center for Materials Research and Technology and Department of 
Physics,\\ 
Florida State University, Tallahassee, FL 32306-4351, USA
\and 
School of Computational Science and Information Technology,\\ 
Florida State University, Tallahassee, FL 32306-4120, USA
\and
Center for Computational Sciences, Oak Ridge National Laboratory,\\
P.O.Box 2008 Mail Stop 6114, Oak Ridge, TN 37831-6114, USA
\and
Department of Physics and Astronomy, Mississippi State University,\\ 
Mississippi State, MS 39762, USA
}

\maketitle              

\begin{abstract}
We discuss numerical and theoretical results for models of magnetization
switching in nanoparticles and ultrathin films. The models and
computational methods include kinetic Ising  and classical Heisenberg 
models of highly anisotropic
magnets which are simulated by dynamic Monte Carlo methods, and
micromagnetics models of continuum-spin systems that are studied by
finite-temperature Langevin simulations. The theoretical analysis builds
on the fact that a magnetic particle or film that is magnetized in a direction
antiparallel to the applied field is in a metastable state. Nucleation
theory is therefore used to analyze magnetization reversal as the decay of
this metastable phase to equilibrium. We present numerical 
results on magnetization reversal in models of nanoparticles and films, 
and on hysteresis in magnets driven by oscillating external fields. 
\end{abstract}

\section{Introduction}

In recent years, the interest in nanostructured magnetic materials has soared 
for a variety of reasons. For one, it is only quite recently that it has 
become possible to synthesize and measure nanometer-sized 
magnetic particles\index{magnetic particles} in small, ordered 
arrays, often by techniques that involve modern, atomic-resolution 
microscopies, such as 
scanning-tunneling microscopy\index{scanning-tunneling microscopy} (STM), 
atomic force microscopy\index{atomic force microscopy} (AFM), 
or magnetic force microscopy\index{magnetic force microscopy} 
(MFM) \cite{DORM92,CROM93,KENT93,KENT94,PAI97}. 
AFM and MFM pictures of an array of 
nanometer-sized iron pillars, fabricated by 
STM-assisted chemical vapor deposition\index{chemical vapor deposition} 
\cite{GIDE96}, are shown in Fig.~\ref{fig:awschalom}.
The techniques are currently 
becoming precise enough to even allow investigation of individual nanoparticles.
At the same time, computers have had a profound 
influence in two different ways. The need for ever higher data-recording 
densities has driven the size of particles used in 
recording media\index{recording media} down into 
the nanometer range \cite{BERT94,CHOU94,NOVO99}, 
while the rapidly increasing power 
of computers has made it feasible to perform 
simulations\index{simulations} of the dynamic 
properties of realistic model systems of sizes comparable to  
experimental ones \cite{RIKV97B,NOWA01,BROW01}. 
\begin{figure}[h,t]
\begin{center}
\end{center}
\begin{center}
\end{center}
\caption[]{
Array of nanoscopic iron pillars of dimensions approximately 
$40 \times 40 \times 200$~nm$^3$, grown by STM-assisted
chemical vapor deposition. 
({\bf a})
AFM image of the array of pillars, as grown on top of a $\umu$m-size 
Hall-effect magnetometer. 
({\bf b})
MFM image of the array after thermal randomization in 
near-zero  applied field. 
The magnetic field from each pillar is imaged and seen to 
point along the major axis of the pillars, 
either up (white) or down (black). 
({\bf c})
MFM image of the array in an applied field of 200~G. Almost 
all the magnets are aligned with the field. 
Image data courtesy of D.D.\ Awschalom and J.~Shi. 
({\bf a}) and ({\bf b}) after \protect\cite{GIDE96}
{\it A pdf version of this paper, which includes this figure, can be 
downloaded from\/} 
http://www/csit.fsu.edu/$\sim{}$rikvold/abstracts/turkey.pdf 
}
\label{fig:awschalom}
\end{figure}

\begin{figure}[h,t]
\begin{center}
\includegraphics[angle=90,width=.8\textwidth]{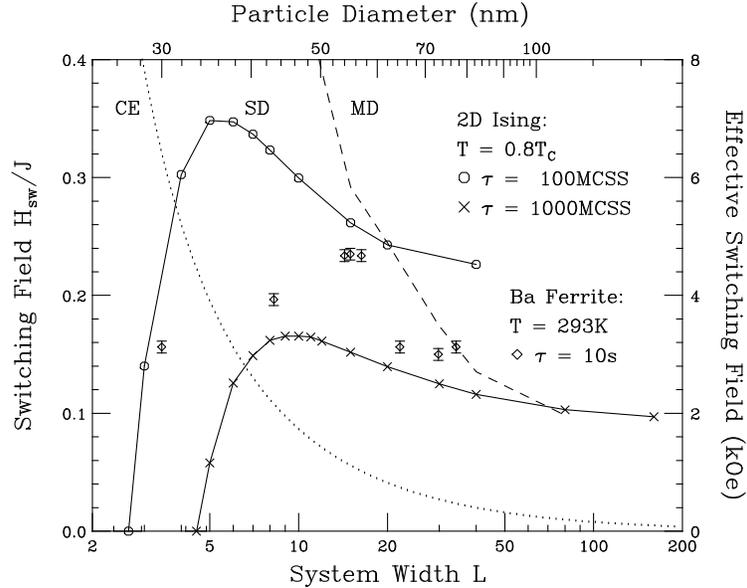}
\end{center}
\caption[]{
Effective switching field (analogous to coercivity) versus particle size 
for single-domain ferromagnetic barium ferrite particles
($\diamond$ with error bars, 
right vertical and top horizontal axes, experimental results) 
and for two-dimensional $L \times L$ Ising systems (data points connected 
by solid lines, left vertical and bottom horizontal axes). 
The barium ferrite data were 
digitized from Fig.~5 of \protect\cite{CHAN93}. 
The Ising data are Monte Carlo simulations from \protect\cite{RICH94} 
for waiting times $\tau$=100 Monte Carlo Steps per Site (MCSS) ($\circ$)
and~1000~MCSS ($\times$) at $T$=0.8$T_{\rm c}$, where $T_{\rm c}$ is the 
exact Ising critical temperature. 
After \protect\cite{RIKV94} 
}
\label{fig:super}
\end{figure}

Modern magnetic recording technologies involve particles that are near
the {\it superparamagnetic limit\/}\index{superparamagnetic limit}. 
In this limit, the energy barrier
separating the two energetically degenerate magnetic orientations is
small enough that thermal fluctuations frequently lead to spontaneous
switching of the orientation. As a result, the magnetic 
coercivity\index{coercivity} decreases 
with decreasing particle size for particles below the superparamagnetic 
size limit. The limit appears as a maximum in a 
curve showing switching field or coercivity versus particle size, such as in 
Fig.~\ref{fig:super}. 
Since the random magnetization reversals\index{magnetization reversal} 
in particles below the 
superparamagnetic limit degrade
recorded information, the engineering challenge has been to keep
the energy barrier in the individual particles high enough to make
spontaneous switching infrequent while keeping the material magnetically 
soft enough
to facilitate recording. As the volumes of the magnetic particles have
shrunk to reach recording densities on the order of
100~Gb/in$^2$ or more  \cite{CHOU94,RUIG00}, materials with higher
coercivities due to strong 
crystalline anisotropies\index{crystalline anisotropy} 
have been employed
\cite{WELLER2000}. In order to enhance engineering practices, it is 
essential to extend the physical understanding of the superparamagnetic limit
past the theories of uniformly mangetized particles to include magnetization
reversal dynamics that proceed through localized regions of reversed 
magnetization that subsequently spread throughout the magnetic element.

The remainder of this article is organized as follows. 
In Sec.~2 we summarize some aspects of the theory of 
magnetization switching\index{magnetization switching} 
in anisotropic magnets\index{anisotropic magnets}, 
including effects of anisotropy
(Sec.~2.1), nucleation theory\index{nucleation theory} 
(Sec.~2.2), and model systems (Sec.~2.3). 
In Sec.~3 we give some new results of finite-temperature
micromagnetic simulations of magnetic nanoparticles. 
In Sec.~4 we discuss hysteresis\index{hysteresis} 
in nanoparticles\index{nanoparticles} 
and ultrathin films\index{ultrathin films}, in 
particular the frequency dependence of hysteresis loops (Sec.~4.1) and a
dynamic order-disorder phase transition\index{dynamic phase transition} 
(Sec.~4.2). A brief summary and 
conclusions are given in Sec.~5. 

\section{Theory of Magnetization Switching in Anisotropic Magnets} 

\subsection{Effects of Magnetic Anisotropy}
 
The most common description of magnetization switching is the
mean-field, uniform-rotation theory of N{\'e}el \cite{NEEL49} and
Brown \cite{BROW59,BROW63}. One assumes uniform rotation of all
localized moments in the particle to avoid an energy barrier due to
exchange interactions of strength $J$. The remaining barrier,
$\Delta$, is caused by magnetic anisotropy -- a combination of
crystal-field and magnetostatic effects.  The equilibrium thickness of
a wall separating oppositely magnetized domains is $\ell_{\rm w} \propto
\sqrt{J/\Delta}$. For particles smaller than $\ell_{\rm w}$ with small
anisotropy, the uniform-rotation picture is reasonable.  If the
anisotropy is largely magnetostatic, the competition between exchange
interactions and the demagnetizing field favors domains of opposite
magnetization in particles larger than $\ell_{\rm w}$. The domains control
switching through the field-driven motion of preexisting domain walls
\cite{KOLE97B,KOLE98C,RIKV00B}.  However, if the anisotropy is largely
crystalline, there exists a range of single-domain particle sizes that
are larger than $\ell_{\rm w}$ but smaller than the size at which the
particle becomes multidomain (often the case in ultrathin films
\cite{HUG96}). 

In anisotropic nanomagnets the state of uniform magnetization opposite
to the applied field constitutes a 
{\em metastable phase\/}\index{metastable phase}.
This nonequilibrium phase decays by thermally
assisted nucleation and subsequent
growth of localized regions, inside which the magnetization is
parallel with the field \cite{RICH94}.  
These growing regions are referred to as {\em droplets}\index{droplets} to
distinguish them from equilibrium domains. 
This mechanism yields results
very similar to recent experiments on single-domain nanoscale
ferromagnets \cite{KOCH}.

\subsection{Application of Nucleation Theory to Magnetization Reversal} 

Here we present a short summary of 
homogeneous nucleation theory as it 
applies to uniaxial magnets\index{uniaxial magnets}. This theory 
covers situations in which the switching 
events are nucleated by thermal fluctuations, 
without the influence of defects. Further details are available in
\cite{RIKV97B,RICH94,RIKV94,RICH95C,RICH96,TOMI92A,RIKV94A}.

The central problems in nucleation theory are to identify the 
fluctuations that lead to the decay of the metastable phase and to 
obtain their free-energy cost relative to the metastable phase. 
For anisotropic systems dominated by short-range interactions, 
these fluctuations are compact droplets of radius $R$.  
The free energy of the droplet has two competing 
terms: a positive surface term $\propto \sigma(T) R^{d-1}$ 
and a negative bulk term $\propto |H| R^d$ 
where $d$ is the spatial dimension, 
$\sigma(T)$ is the surface tension of the droplet wall, and $H$ is the applied 
magnetic field along the easy axis.  
Their competition yields a critical droplet radius, 
$ R_c(H,T) \propto {\sigma(T)}/{|H|}$.  
Droplets with $R<R_c$ 
most likely decay, whereas droplets with $R>R_c$ most likely grow 
to complete the switching process. 
The free-energy cost of the critical droplet ($R=R_c$) is 
$\Delta F(H,T) \propto \sigma(T)^d / |H|^{d-1}$. 
Nucleation of critical droplets at nonzero temperature $T$ 
is a stochastic process with nucleation rate per unit volume 
given by an Arrhenius relation: 
\begin{equation} \label{eq:NucRate}
I(H,T) \propto 
|H|^{K} \exp \left[ - { \beta \Delta F(H,T) } \right] 
\equiv 
|H|^{K} \exp \left[ - {\beta \Xi(T) / |H|^{d-1}} \right] 
\; ,
\end{equation}
where $\beta = 1/ k_{\rm B}T$ ($k_{\rm B}$ is Boltzmann's constant), 
$\Xi(T)$ is the $H$-independent part of $\Delta F$, and 
the prefactor exponent $K$ is known for many models 
from field-theoretical arguments  \cite{RIKV94,LANG67,LANG69,GNW80}.
The particles are of finite size $L$, and the dominant reversal mechanism 
depends on $H$, $T$, and $L$.
\begin{figure}[t]
\includegraphics[width=1.0\textwidth]{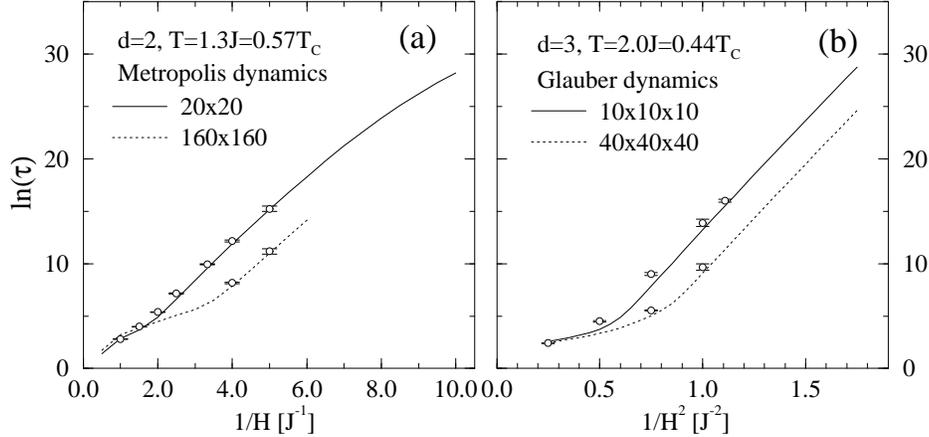} 
\caption[]{
Lifetimes for two-dimensional $L^2$ Ising systems with $L = 20$
and~160 at $T = 0.57T_c$ ({\bf a}), and three-dimensional $L^3$
Ising systems with $L=10$ and~40 at $T=0.44T_c$ ({\bf b}).  The
data points are direct Monte Carlo simulation results, while the lines
are extrapolations with the Projected Dynamics (PD) accelerated
dynamics algorithm \cite{KOLE98B,KOLE98A,NOVO99B}, based on the
smallest system at the weakest field.  The sharp changes in slope
correspond to the DSp, with the deterministic regime to the left and
the stochastic regime to the right.  
The ratio of the slopes of the curve in the 
single-droplet and multidroplet regimes is $(d+1)$, 
in agreement with (\ref{eq:tauSD}) and (\ref{eq:tauMD}).
After \protect\cite{KOLE98A}
}
\label{fig:tauvsh}
\end{figure}

In the weakest applied fields, the particles are in the
``Coexistence'' (CE) regime, with the average metastable lifetime 
$\tau_{\rm CE}(H,T,L) \sim \exp \left[ {2 \beta \sigma (T) L^{d-1}}
\right]$. (This result is nearly independent of the boundary
conditions \cite{RICH96}.) The regime corresponds to $R_{\rm c} > L,$
and the associated $L$-dependent crossover field is called the
Thermodynamic Spinodal (ThSp)
\cite{RIKV94,TOMI92A,RIKV94A}. Estimating its value by assuming $R_{\rm
c}(H,T,L) \approx L,$ one finds $H_{\rm ThSp}(T,L) \sim L^{-1}$. 
The $L$-dependence of $H_{\rm ThSp}$ is given by the dotted curve in 
Fig.~\ref{fig:super}.

{}For $|H| > H_{\rm ThSp}$ (but not too large), 
the lifetime is determined by the inverse of the total nucleation rate, 
\begin{equation}
\label{eq:tauSD}
\tau_{\rm SD}(H,T,L) \approx \left( L^d I(H,T) \right)^{-1} 
\propto 
L^{-d} |H|^{K} \exp \left[ {\beta \Xi(T) |H|^{d-1}} \right] 
\; ,
\end{equation}
which is inversely proportional to the particle volume, $L^d$ (see
Fig.~\ref{fig:tauvsh}).  The subscript SD stands for Single Droplet
and indicates that in this regime the switching is normally completed
by the first droplet to reach $R_c$.  In both of the stochastic reversal
regimes(CE and SD) the probability that switching has not taken
place within a time $t$ after the field reversal, $P_{\rm not}(t)$,
takes the form $P_{\rm not}(t) = \exp(-t/\tau)$.

A second crossover, called the Dynamic Spinodal (DSp)
\cite{RIKV94,TOMI92A,RIKV94A}, is a consequence of the finite
velocity, $v \approx \nu |H|$, of the surface of a growing
supercritical droplet \cite{RIKV00B}.  A reasonable criterion to locate
the DSp is that the average time between nucleation events, which is
$\tau_{\rm SD}$, should equal the time it takes a droplet to grow to a
size comparable to $L$. This yields the asymptotic relation $H_{\rm
DSp}(T,L) \sim [\ln(L)]^{1/(d-1)}$.  
The $L$-dependence of $H_{\rm DSp}$ is given by the dashed curve in 
Fig.~\ref{fig:super}. 
{}For $|H| > H_{\rm DSp}$, the
metastable phase decays through many droplets which nucleate and grow
independently in different parts of the system.  In this Multidroplet
(MD) regime \cite{RIKV94,TOMI92A,RIKV94A}, the classical
Kolmogorov-Johnson-Mehl-Avrami (KJMA) theory of metastable decay in
large systems \cite{KOLM37,JOHN39,AVRAMI} gives the lifetime
\begin{equation}
\label{eq:tauMD}
\tau_{\rm MD}(H,T) 
\propto \left[ {I(H,T) (\nu |H|)^d}/{(d+1) \ln 2}
            \right]^{- {1}/{(d+1)}}
\;,
\end{equation}
{\em independent\/} of $L$ (see Fig.~\ref{fig:tauvsh}). 
In the MD regime 
$P_{\rm not}(t)\approx {\rm erfc}\left( (t-\tau)/\Delta \right)$ 
\cite{RICH94}, where the width $\Delta$ of the switching-time distribution 
depends on $H$, $T$, and $L$. 
\begin{figure}[t]
\sidecaption
\includegraphics[width=0.60\textwidth]{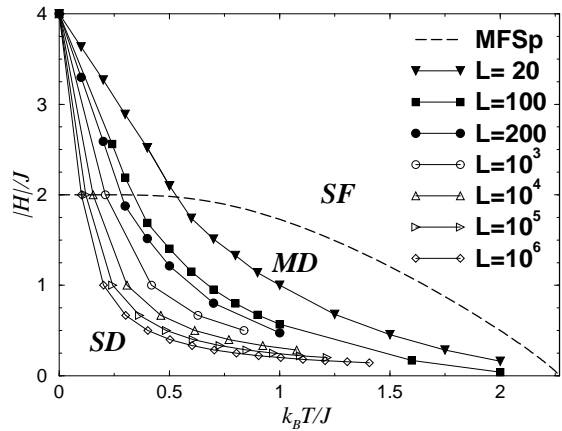} 
\caption[]{
Metastable phase diagram for the two-dimensional Ising model at temperatures 
up to $T_c$. The dashed curve represents $H_{\rm MFSp}(T)$. Data points 
connected by solid line segments represent $H_{\rm DSp}(T)$ for several values 
of $L$ between 20 and 10$^6$. The filled data 
points are the results of Monte Carlo 
simulations, while the empty data points represent a 
low-temperature approximation \cite{JLEE94A} for large systems 
}
\label{fig:metapd}
\end{figure}

For very strong fields nucleation theory becomes irrelevant to the switching 
behavior. A reasonable way to estimate the crossover to this Strong-Field (SF) 
regime is to require that the critical radius should be on the order of the 
lattice constant $a$. Specifically, requiring $R_c = a/2$, we get the 
crossover field called the mean-field spinodal (MFSp)
$H_{\rm MFSp}(T) \approx 2 \sigma(T) / m_{\rm eq}(T)$, 
where $m_{\rm eq}(T)$ is the 
zero-field equilibrium magnetization. The ``metastable phase diagram'' in 
Fig.~\ref{fig:metapd} shows 
$H_{\rm MFSp}(T)$, as well as $H_{\rm DSp}(T)$ for two-dimensional Ising 
systems of widely varying 
sizes. Note the logarithmically slow convergence to zero of 
$H_{\rm DSp}$ with increasing system size. As a result, even macroscopic 
metastable systems may be ``small'' 
in the sense that they decay via the single-droplet mechanism. 

The switching field, $H_{\rm sw}(t_{\rm w},T,L)$, is the field 
required to observe a specified average waiting time, $t_{\rm w}$. It 
is found by solving $\tau$ in the relevant region (CE, SD, or MD) 
for $H$ with $\tau$$=$$t_{\rm w}$.  
The resulting $L$ dependence of $H_{\rm sw}$ is 
a steep increase with $L$ in the CE regime, 
peaking near the ThSp, followed by a decrease in the SD regime towards a 
plateau in the MD regime \cite{RICH94,RICH95C}. 
This behavior is illustrated in Fig.~\ref{fig:super}. 
Note that the maximum in $H_{\rm sw}$ (related to the maximum coercivity) 
occurs {\em even in the absence of dipole-dipole interactions}.  
{}For other boundary conditions 
and in systems with dipole interactions the $H_{\rm sw}$ versus $L$ curve can 
even have more that one maximum \cite{RICH96}.  
Maximizing the coercive field is important in magnetic recording 
applications.

\subsection{Statistical-mechanical Model Systems}

Simplified statistical-mechanical models are often 
amenable to analytic solutions, with no free fitting parameters, that
agree well with the numerical results. Despite their lack of realism, they 
are therefore important as testing grounds for theoretical descriptions of 
different switching mechanisms. In more realistic models,
that correspond to larger numbers of actual materials, analytic
results are difficult to come by, but the physical insights gained from
the simpler models can be readily applied. Here we introduce three such 
models in order of 
increasing complexity: the kinetic Ising model, the classical Heisenberg model, 
and finite-temperature Langevin micromagnetic models. 

\subsubsection{The Ising Model}

The simplest microscopic model of a ferromagnet is the nearest-neighbor 
Ising model\index{Ising model}, in which 
discrete spins, $s_i = \pm 1$, are placed on the sites (labeled $i$) of a 
two- or three-dimensional lattice. The spins interact with their neighbors 
with a strength $J$, so that the model is described by the Hamiltonian
\begin{equation}
{\mathcal H} = - J \sum_{\langle i,j \rangle} s_i s_j - H \sum_i s_i \;.
\label{eq:Ising}
\end{equation}
The model can easily be generalized to longer-range interactions, different 
lattice geometries, etc. Despite its apparent simplicity, it has many of the 
attributes of more complicated systems, while many of its 
properties are exactly known. It is therefore a very commonly studied model. 

The Ising Hamiltonian, (\ref{eq:Ising}), is not a true quantum-mechanical 
Hamiltonian, and the Ising model therefore does not have an intrinsic dynamic. 
To simulate thermal fluctuations one uses 
Monte Carlo simulation\index{Monte Carlo simulation} of 
a local stochastic dynamic which does not conserve the order parameter. 
An often-used example is the Metropolis \cite{METR53} 
dynamic\index{Metropolis dynamic} with the spin-flip probability 
\begin{equation}
W_{\rm M}(\beta \Delta E) 
= \min \left[ 1, \exp \left(- \beta \Delta E \right) \right] 
\;,
\label{eq:Metro} 
\end{equation}
where $\Delta E$ is the energy change that 
would ensue if the flip were to occur. 
Another popular choice is the Glauber dynamic\index{Glauber dynamic} 
\cite{GLAU63}, defined by 
\begin{equation}
W_{\rm G}(\beta \Delta E) = \frac{\exp \left(- \beta \Delta E \right)}
{1+ \exp \left(- \beta \Delta E \right)} \;.
\label{eq:Glau}
\end{equation}
The basic time scale of the Monte Carlo simulation is not known from first
principles, but it is expected to be on the order of a typical inverse
phonon frequency, 10$^{-9}$--10$^{-13}$~s. In dynamics such as these,
where each potential flip is accepted or rejected randomly, flips can
become very rare when rejection rates are high.  To perform
simulations on the very long time scales necessary to observe metastable
decay, one needs to use rejection-free Monte Carlo algorithms
\cite{NOVO95,BORT75,GILM76,NOVO95A} and other advanced algorithms 
\cite{KOLE98B,KOLE98A,NOVO99B}.

Analogous stochastic time evolutions can also be imposed on models whose
spins have continuous degrees of freedom.  Here we briefly discuss one
such model,  the anisotropic Heisenberg model.

\subsubsection{The Anisotropic Heisenberg Model} 

Like the Ising model, 
the Heisenberg model\index{Heisenberg model} 
consists of spins located at discrete points on a lattice.
However, unlike the spins in the Ising model, which equal $\pm 1$,
Heisenberg spins are $n$-dimensional vectors of unit length.
When $n=2$, this model is usually referred to as the $XY$ or plane-rotor model.
The Hamiltonian for the nearest-neighbor 
Heisenberg model with only interaction anisotropy is
\begin{equation}
{\mathcal H} = - \sum_{\langle i,j \rangle} \sum_n 
\left[ J_n s_{n, i} s_{n, j} \right] - \vec{H} \cdot \sum_i \vec{s}_i \;,
\label{eq:Heis}
\end{equation}
where $s_{n,i}$ is the $n$-th component of the $i$-th spin vector $\vec{s}_i$, 
$J_n$ are coupling constants, and $\vec{H}$ is the external magnetic field.
For the example of this model presented here, 
$n=3$, $J_x=J_y=1$, $J_z=2$, $\vec{H}=H_z \hat{z}$, 
and the lattice is a two-dimensional $L \times L$ square lattice.

There are many stochastic dynamics for the Heisenberg model which yield 
identical equilibrium results, but have different relaxation dynamics.
The dynamic assumed here consists of selecting a spin vector at random,
then choosing a new orientation for that spin, uniformly distributed 
over the unit sphere, and then accepting or 
rejecting the new configuration based on (\ref{eq:Glau}).
Other dynamics exist which make only small changes to spin orientations,
but these are not discussed here.
\begin{figure}[h,t]
\begin{center}
\includegraphics[angle=00,width=.85\textwidth]{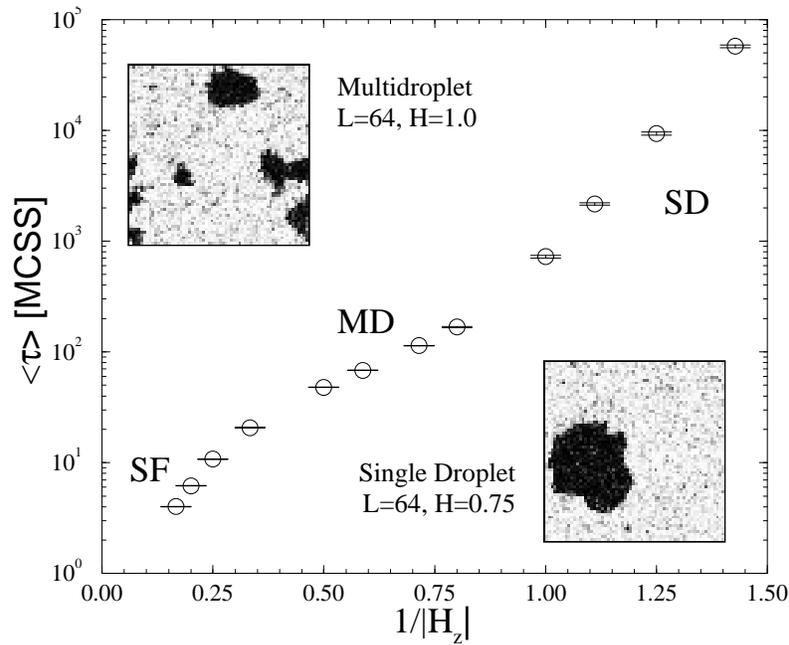}
\end{center}
\caption[]{ 
Switching behavior for an anisotropic $n=3$ Heisenberg model 
on a two-dimensional $L \times L$ square lattice.
The parameters are $J_x=J_y=1$, $J_z=2$, and $T=1$, 
which is below the critical temperature for this model.
The circles are for a system with $L=16$, and 
the single-droplet (SD), multidroplet (MD), and strong-field (SF) 
regimes are labeled.
The insets show typical system configurations during the switching 
process for systems with $L=64$.
Note that the dynamic spinodal (DSp) depends on $L$,  
and thus the SD and MD regimes appear at different fields for $L=16$ and $L=64$.
The grayscale for the insets shows $s_z$ with lighter shades indicating 
metastable spins and darker shades indicating more stable spins
}
\label{fig:heisen}
\end{figure}

The simulation begins with all $\vec{s}_i=-\hat{\vec z}$ and $H_z > 0$.
This metastable phase 
then decays in a manner consistent with homogeneous nucleation and growth.
However, unlike the Ising model, 
the continuous degrees of freedom add additional complications,
such as effective long-range interactions between droplets.
We have not yet attempted to quantify these differences.
Figure~\ref{fig:heisen} shows lifetimes and configuration snapshots in
the single-droplet, multidroplet, and strong-field regimes
for an anisotropic Heisenberg model at a temperature below criticality. 
The field dependence of the lifetime is seen to be very similar to that
of the two-dimensional Ising model, shown in 
Fig.~\ref{fig:tauvsh}({\bf a}). 

\subsubsection{Finite-temperature Langevin Micromagnetics}

\begin{figure}[h,t]
\sidecaption
\includegraphics[angle=00,width=.60\textwidth]{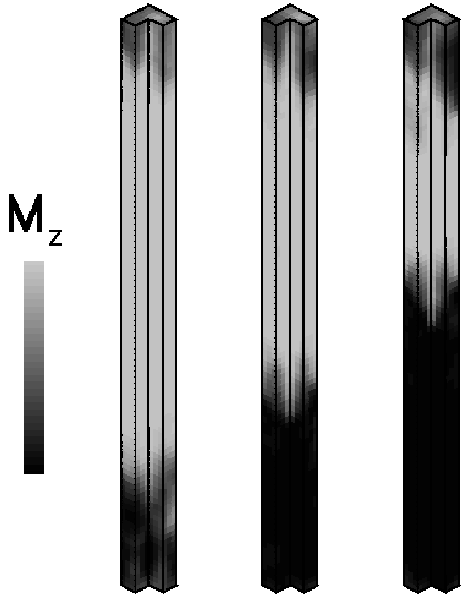}
\caption[]{ 
Magnetization along the pillar long axis, $M_z$, at three times 
during the switching process. Light shades represent the metastable
orientation and dark shades the equilibrium orientation. The
micromagnetic simulation shown in this figure  models an individual pillar 
using a $7$ $\times$ $7$ $\times$ $49$ lattice at $T=20$~K. 
The pillar is shown in a one-quarter cut-away view
}
\label{fig:pillar}
\end{figure}

More realistic representations of nanoscale magnetic systems can be
obtained by micromagnetic modeling. In this method the ``spins'' are
coarse-grained magnetization vectors ${\vec{M}}({\vec{r}}_i)$; each
represents the magnetization within a cell centered at
position ${\vec{r}_i}.$ In this low-temperature model \cite{GARA97},
the vectors have a fixed magnitude $M_s$ corresponding to the bulk
saturation magnetization density.  The time evolution of each spin is
governed by the damped precessional motion given by the
Landau-Lifshitz-Gilbert (LLG) equation \cite{BROW63B,AHAI96}
\begin{equation}
\label{eq:llg}
\frac{ {\rm d} {\vec{M}}({\vec{r}}_i) }
     { {\rm d} {t} }
 =
   \frac{ \gamma_0 }
        { 1+\alpha^2 }
   {\vec{M}}({\vec{r}}_i)
 \times
 \left (
   {\vec{H}}({\vec{r}}_i)
  -\frac{\alpha}{M_s} {\vec{M}}({\vec{r}}_i) \times 
                      {\vec{H}}({\vec{r}}_i)
 \right )
\;,
\end{equation}
where the electron gyromagnetic ratio is $\gamma_0 = 1.76 \times 10^7
\;{\rm Hz/Oe}$ \cite{AHAI96}, and $\alpha$ is a phenomenological damping
parameter. The local field at the $i$-th spin,
${\vec{H}}({\vec{r}}_i)$, is generally different at each location.
This field is a linear superposition of fields, one for
each type of interaction in the system. Typical examples include
fields from external sources, exchange, crystalline anisotropy, and
dipole-dipole interactions. Thermal fluctuations may also contribute a
term: a stochastic field ${\vec{H}}_{\rm
n}({\vec{r}}_i)$ that is assumed to fluctuate independently for each
spin \cite{BROW63}. The fluctuations are assumed Gaussian, each with
zero first moment and with the second moments given by the
fluctuation-dissipation relation \cite{BROW63}
\begin{equation}
\label{eq:fluctuation}
\langle
 {{H}}_{{\rm n} \mu}({\vec{r}}_i,{t}) 
 {{H}}_{{\rm n} \mu'}({\vec{r}}_i',{t}')
\rangle
=   \frac{2 \alpha k_{\rm B}T}{\gamma_0 M_s V}
    \delta\left({t}-{t}'\right)
    \delta_{\mu,\mu'}
    \delta_{i,i'}
\;,
\end{equation}
where ${{H}}_{{\rm n} \mu}$ indicates one of the Cartesian components
of $\vec{H}_{\rm n}$. Here $V=(\Delta {r})^3$ is the discretization
volume of the cell, $\delta_{\mu,\mu'}$ is the Kronecker delta
representing the orthogonality of the Cartesian coordinates, and
$\delta({t}-{t}')$ is the Dirac delta function.

While this stochastic term necessitates careful treatment of the
numerical integration in time of this stochastic differential 
(Langevin) equation\index{Langevin equation}, 
the most computationally intensive part
of the calculation involves the dipole-dipole term. For systems with
more than a few hundred model spins, it is necessary to use a
sophisticated algorithm such as the 
fast multipole method\index{fast multipole method} (FMM)
\cite{GREE87A,GREE94}. An extensive discussion of the issues involved
in finite-temperature simulations of micromagnetics is presented in
\cite{BROW01}. The growth of a droplet during the switching of
an iron nanopillar at $T$$=$$20\,{\rm K}$ is shown in
Fig.~\ref{fig:pillar}. Nucleation is observed to occur
at the ends of the pillars \cite{BROW01}.

\section{Finite-temperature Micromagnetics Results for Nanoparticles}

Micromagnetic simulations have been applied to the study of iron
pillars modeled after those shown in Fig.~\ref{fig:awschalom}. 
Unless otherwise noted, 
the model iron pillars discussed here are $5.2\,{\rm nm}$ $\times$
$5.2\,{\rm nm}$ $\times$ $88.4\,{\rm nm}$. 
The cross-sectional dimensions are small enough, about two exchange
lengths, that the only significant inhomogeneities in the
magnetization are those in the $z$ direction, {\em i.e.\/} along
the long axis of the pillar \cite{HINZKE,NOWAK2}. 
\begin{figure}[b]
\sidecaption
\includegraphics[angle=00,width=.65\textwidth]{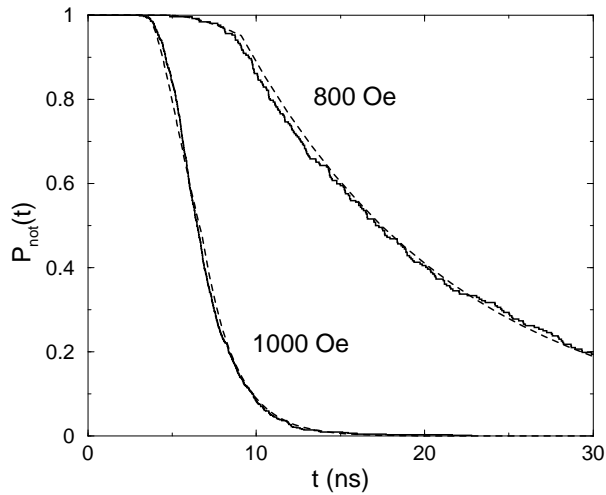}
\caption[]{ 
Probability of not switching before time $t$, $P_{\rm
not}(t)$, for micromagnetic simulations at $T$$=$$100\,{\rm K}$ with
applied fields of $1000\,{\rm Oe}$ and $800\,{\rm Oe}$ as labeled.
The solid curves are simulation data for $624$ and $252$ switches,
respectively, for pillars modeled as a one-dimensional chain of
spins. The dashed curves are fits to the theoretical model, (\ref{eq:twoexpo})
}
\label{fig:twoexpo}
\end{figure}
In light of this, the
pillars have been modeled as a linear system of magnetic cubes with
side $5.2\,{\rm nm}$.  This model, discussed previously in
\cite{BROW01,BOER97}, includes thermal fluctuations, exchange, and
dipole-dipole interactions.

The results for $P_{\rm not}(t)$, with $T$$=$$100\,{\rm K}$ are shown
in Fig.~\ref{fig:twoexpo} for applied fields of
$H$$=$$1000\,{\rm Oe}$ and $800\,{\rm Oe}$. Here switching is defined
to occur when the $z$-component of the total magnetization,
$M_z$, passes through zero. The form of $P_{\rm not}(t)$ is not
exponential, which can be explained by the fact that nucleation of the
reversed droplets is easier at the ends of the pillars than in the middle. 
Assuming that the nucleation rate for droplets at the end of
the pillars is constant, $I$, and that the earliest time switching can
occur because of the finite velocity of
droplet growth is $t_0,$ it can be shown that the
probability of not switching is \cite{BROW01}
\begin{equation}
\label{eq:twoexpo}
P_{\rm not}\left(t\right) = 
\left\{
\begin{array}{lr}
  1                               \qquad\qquad & t< t_0 \\
  e^{-2{I}(t-t_0)}             
    \left[1+2{I}(t-t_0)\right] \qquad\qquad & t_0 \le t< 2t_0 \\
  e^{-2{I}(t-t_0)}             
    \left[1+2{I}t_0\right]     \qquad\qquad & {2t_0\le t}
\end{array}
\right.
\;.
\end{equation}
The parameters are fitted by matching the first and
second moments of the simulation results to those of the theoretical
forms. As long as the applied field is relatively weak, the agreement
between the theory and the simulations is quite good. Switching at
times $t_0 \le t < 2t_0$ is possible only when nucleation occurs at both ends. 

\begin{figure}[b]
\sidecaption
\includegraphics[angle=00,width=.65\textwidth]{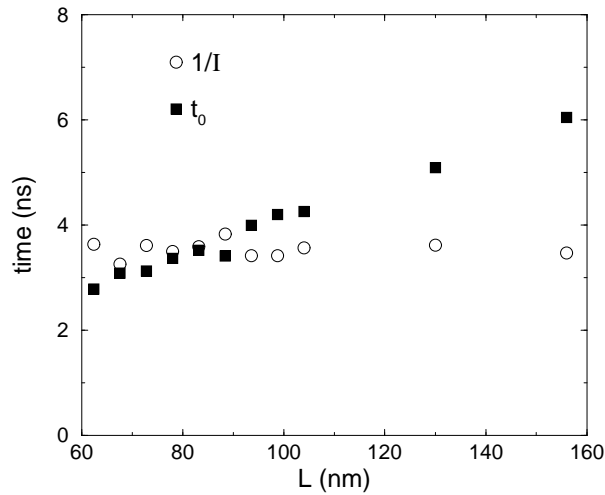}
\caption[]{
Inverse nucleation rate, $I^{-1}$, and earliest switching time, $t_0$, from
(\protect\ref{eq:twoexpo}), for different pillar lengths. The rate of
nucleation at the ends of the pillars depends only weakly on the
pillar length, and therefore likewise the energy barrier. The earliest
switching time increases since the droplets have to grow for a longer time
to switch longer pillars  
}
\label{fig:mmsize}
\end{figure}

Since the nucleation occurs at the ends, the dependence of the
switching on the size of the system is different from that seen in 
isotropic models. Results for the parameters $t_0$ (squares) and
$1/I$ (circles) at $T$$=$$100\,{\rm K}$ and $H$$=$$1000\,{\rm Oe}$ are
shown in Fig.~\ref{fig:mmsize} for pillars of different lengths, {\em
i.e.\/} composed of different numbers of cubes $5.2\,{\rm nm}$ on a
side. The nucleation rate is nearly constant, indicating that the size
of the energy barrier does not depend on the pillar length. The growth
time, indicated by $t_0$, however, increases as the droplets have to
grow farther to switch the magnetization. The nearly linear increase
with pillar length indicates that the interface velocity is not
significantly affected by the demagnetizing field associated with the
high aspect ratio of the pillars.

\begin{figure}[t]
\sidecaption
\includegraphics[angle=00,width=.65\textwidth]{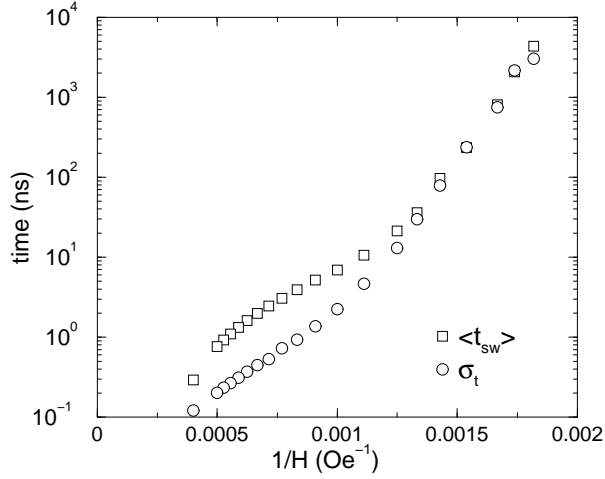}
\caption[]{
For very weak fields, $\langle
t_{\rm sw}\rangle$ and $\sigma_{t}$ are equal, 
indicating an exponential probability density for the switching times
in a single-droplet mode. At higher fields $\sigma_{t}$
decreases faster as the switching process becomes nearly
deterministic. This is associated with the multidroplet switching mode
}
\label{fig:mmswitch}
\end{figure}

Finally, changes in the switching mode as the field is changed are
shown in Fig.~\ref{fig:mmswitch}. Here the mean switching time,
$\langle t_{\rm sw} \rangle$, and standard deviation, $\sigma_t$, are
shown versus applied field for the $88.4\,{\rm nm}$ long pillars at
$T$$=$$100\,{\rm K}.$ 
The mean, $\langle t_{\rm sw} \rangle,$ and standard deviation,
$\sigma_{t},$ of the switching time $t_{\rm sw}$ versus inverse 
applied field for pillars of the same type as considered in  
Fig.~{\ref{fig:twoexpo}}. 
At weak fields the mean and standard deviation
are nearly equal, where the exponential tail at $t>2t_0$ dominates
(\ref{eq:twoexpo}). As the applied field is increased, the barrier
to nucleation decreases, and the exponential behavior becomes less
dominant. Eventually, (\ref{eq:twoexpo}) breaks down as
the multidroplet reversal mechanism becomes important. The
multidroplet nature of the reversal has been verified by direct
observation of the switching, which shows droplets nucleating away
from the ends at $H$$=$$1000\,{\rm Oe}.$

\section{Hysteresis}

Hysteresis is common in many nonlinear systems driven by an 
oscillating external force, including nanostructured magnets in an oscillating 
field. It occurs when the dynamics of the system is too sluggish to keep 
pace with the force. The term was coined by Ewing in the 
context of magnetoelasticity \cite{EWIN1881} 
from the Greek word  {\it husterein\/}
\mbox{($\stackrel{\scriptscriptstyle c}{\upsilon} \! \! \sigma \tau
\epsilon \rho \acute{\epsilon} \omega$)}
which means ``to be behind.'' 

\subsection{Hysteresis-loop Areas}

\begin{figure}[h,t]
\sidecaption
\includegraphics[angle=0,width=.7\textwidth]{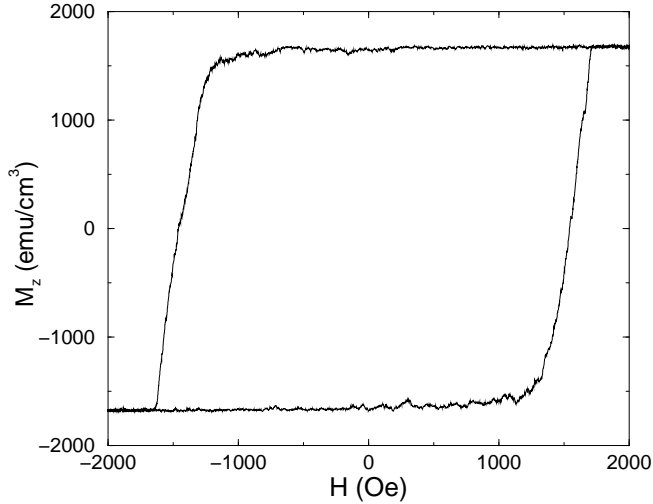}
\caption[]{
Hysteresis loop for one model Fe pillar of the same type considered in  
Fig.~\protect{\ref{fig:twoexpo}}
at $T=100$~K for an oscillating field with a period of $75\,{\rm ns}$ 
}
\label{fig:loop}
\end{figure}
Among the earliest aspects of hysteresis to receive sustained interest is 
the hysteresis-loop area. In the magnetic context of this article, the 
hysteresis loop is a plot of magnetization versus applied field, and its area 
is given by the integral $A = - \oint m(H) {\rm d} H$. A typical 
hysteresis loop for a small system 
with thermal noise is shown in Fig.~\ref{fig:loop}.  
\begin{figure}[h,t]
\begin{center}
\includegraphics[angle=270,width=.65\textwidth]{fig11a_aMDlowfreq-fig9.epsi}
\includegraphics[angle=00,width=.65\textwidth]{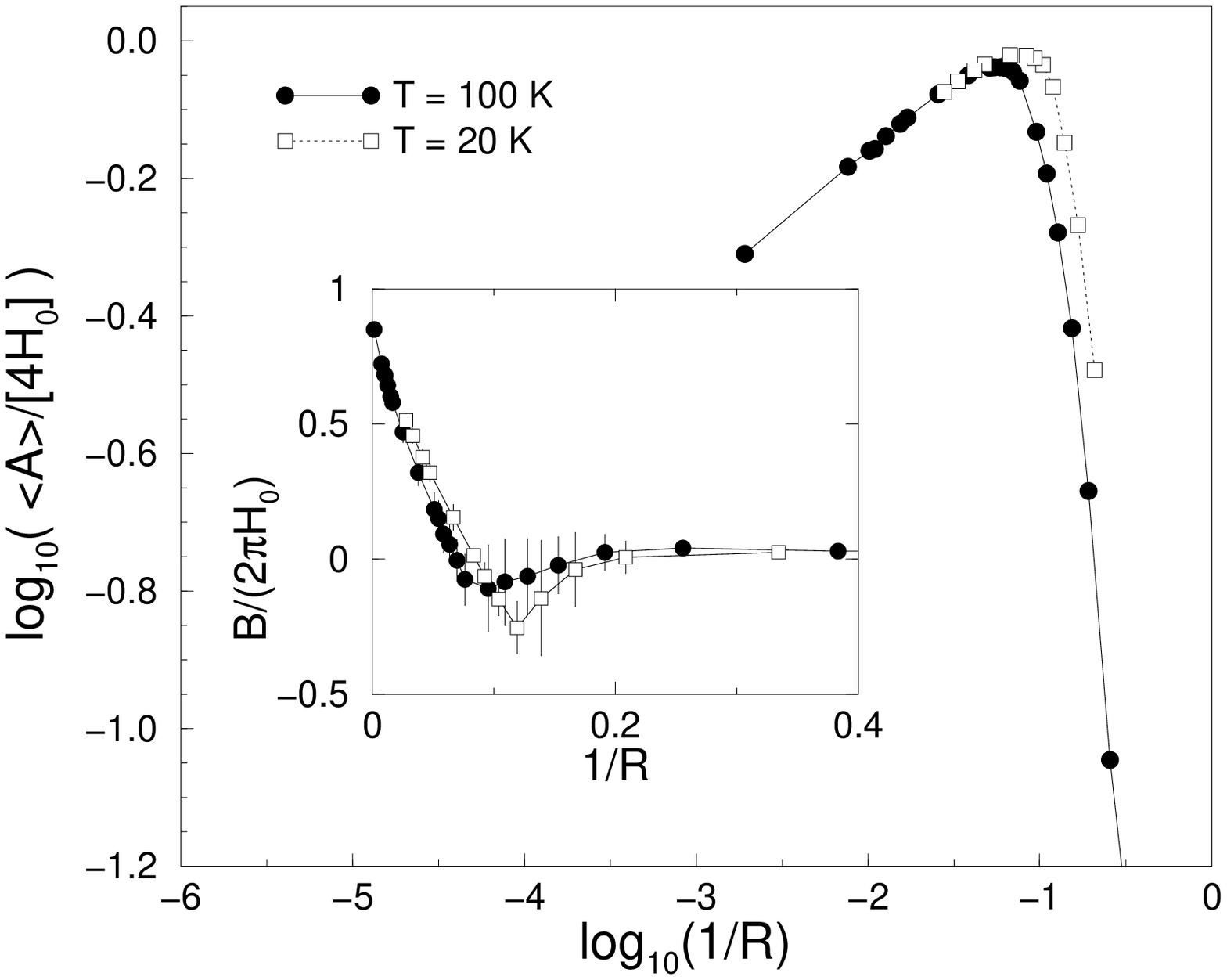}
\end{center}
\caption[]{
The hysteresis-loop area, $A$, versus the dimensionless frequency, $1/R$. 
({\bf Top}) 
Two-dimensional Ising model of an ultrathin film. 
Data points: Monte Carlo simulations for $L=64$, $T$$=$$0.8T_c$, 
and $H_0$$=$$0.3J$. 
For these parameters the magnetization switching occurs 
via the multidroplet mechanism, except for the lowest frequencies. 
Solid curve: numerical integration for sinusoidally varying field. 
Dotted curve: numerical integration for linearly varying field. 
Dot-dashed curve: numerical integration assuming magnetization reversal via 
the single-droplet mechanism. 
Dashed curve: low-frequency asymptotic solution. 
Power-law fits would yield 
very different effective exponents for fits centered at different frequencies. 
After \cite{SIDE99}. 
({\bf Bottom}) 
Micromagnetic model of iron pillars with length $88.4\,{\rm nm}$ and
square cross-section $5.2\,{\rm nm}$ $\times$ $5.2\,{\rm nm}$ for
$T$$=$$100\,{\rm K}$ (circles) and $T$$=$$20\,{\rm K}$ (squares) with
$H_0$$=$$2000\,{\rm Oe}.$ The lines are guides to the eye.  The inset
shows the correlation between the magnetization and field versus frequency.
The lowest-frequency zero-crossing, indicating a resonance condition, occurs
at roughly the same frequency as the maximum in $A$
}
\label{fig:loopa}
\end{figure}
The particular importance of the loop area is that it corresponds to the 
energy dissipation per period of the applied field. It is thus relevant to 
the performance of 
most electrical and electronic equipment. Recent experiments on ultrathin 
Fe and Co films with Ising-like anisotropy 
have considered the frequency dependence of the hysteresis-loop areas
\cite{HE93,JIAN95,SUEN97,SUEN99}. The results of these studies 
were interpreted in terms of power laws, but with exponents that vary 
widely between experiments. 
The experimental situation thus may appear somewhat unclear. 

A resolution is provided by the nucleation-and-growth picture of magnetization 
switching presented here. We assume that the system is driven by a sinusoidally 
oscillating field, $H(t) = H_0 \sin (\omega t)$. 
Since the nucleation rate, which has the dimension 
of a frequency per unit volume, is proportional to 
$\exp \left[- \beta \Xi(T) / |H|^{d-1} \right]$ by (\ref{eq:NucRate}), 
one would expect that the 
field at which the magnetization changes sign should depend on the frequency 
as $- \left( \ln \omega \right)^{-1/(d-1)}$. The loop area 
is approximately proportional to 
the switching field multiplied by the saturation magnetization 
(see Fig.~\ref{fig:loop}). 
Thus, one would expect the loop area to show this logarithmic frequency 
dependence in the asymptotic low-frequency limit. Analytic calculations 
have confirmed this asymptotic result. However, 
for higher, but still low, frequencies 
they show a very slow crossover to the asymptotic behavior, 
which is confirmed by Monte Carlo simulations. Such a slow crossover could 
easily be mistaken for a power law, even when observed over several decades in 
frequency \cite{SIDE99,SIDE98B,SIDE98A}. This result was shown to hold, 
both when the magnetization 
reversal occurs via the single-droplet mechanism \cite{SIDE98A} and the 
multidroplet mechanism \cite{SIDE99}, even though the details are different. 
The behavior is illustrated in 
the top part of Fig.~\ref{fig:loopa}. 
Recently we have also found analogous 
behavior in micromagnetics simulations of nanometer-sized iron pillars, 
see the bottom part of Fig.~\ref{fig:loopa}. 
In these figures the frequency is given in terms of the dimensionless frequency 
$1/R = \omega \tau(H_0,T) / 2 \pi$. 

\subsection{Dynamic Phase Transition}

Different phenomena occur in hysteretic systems as the driving frequency is 
increased. Eventually the field will vary too quickly for the system to have 
time to switch during a single period. 
\begin{figure}[h,t]
\includegraphics[angle=00,width=.5\textwidth]{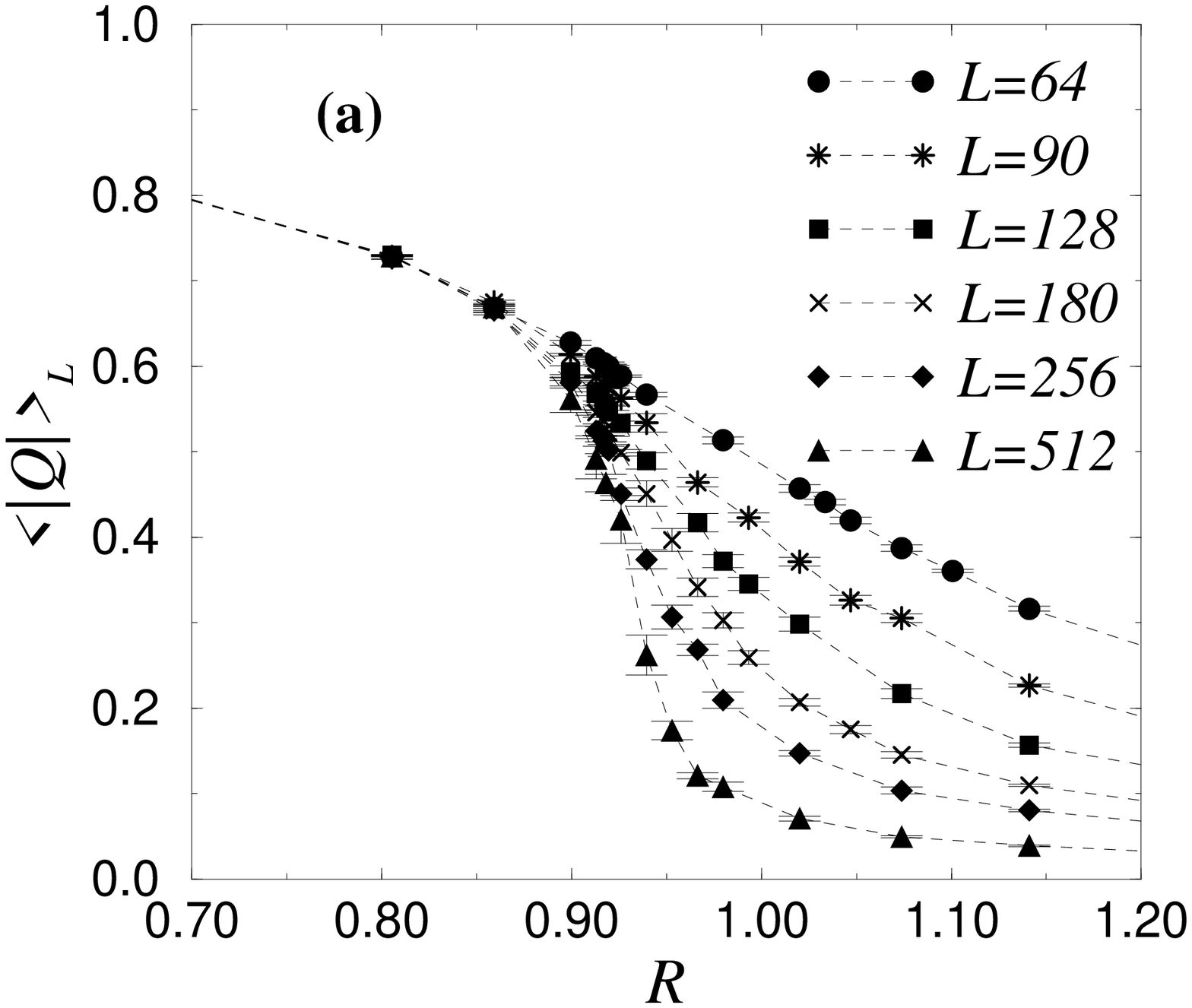}
\includegraphics[angle=00,width=.5\textwidth]{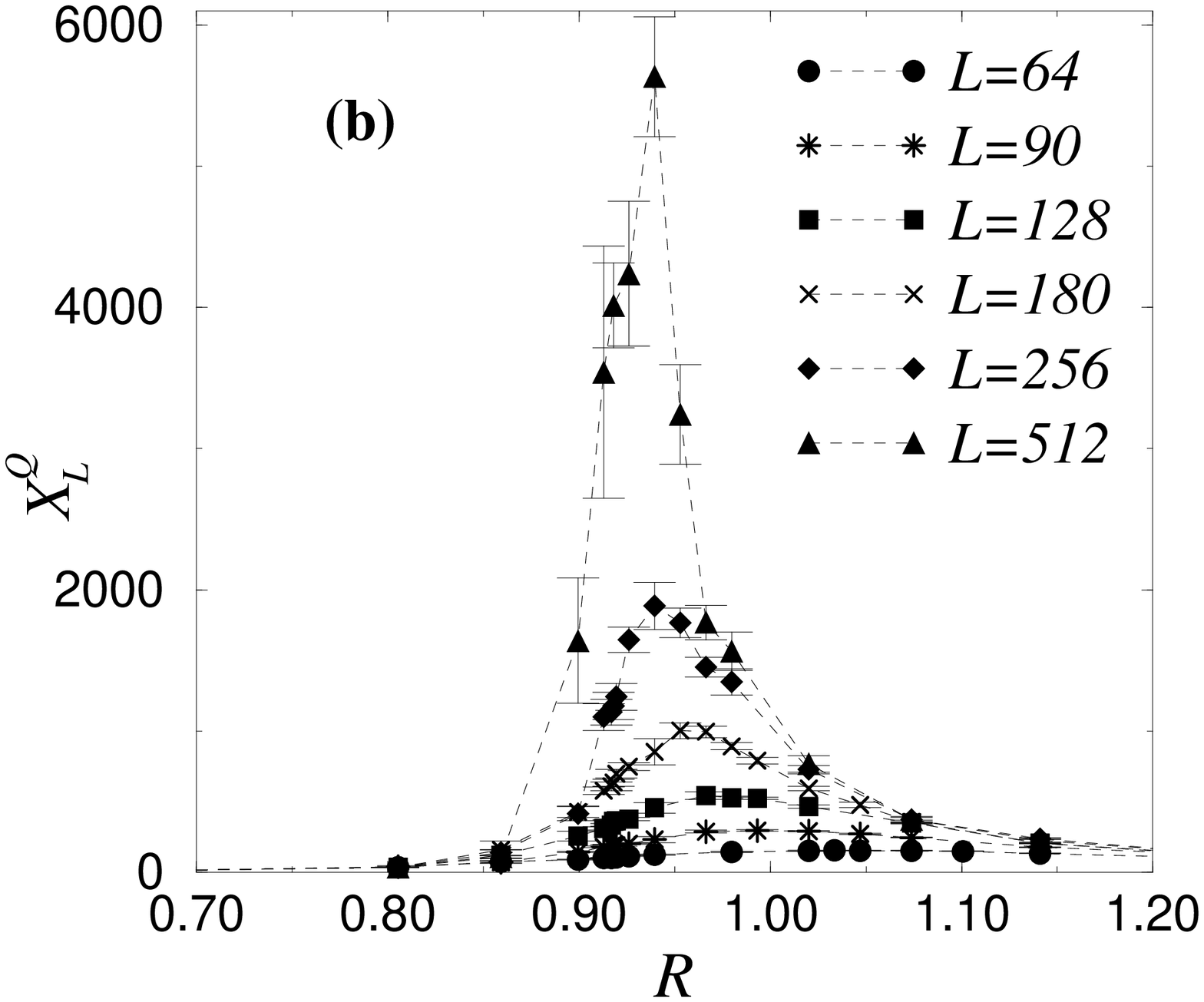}
\caption[]{
The dynamic phase transition in a two-dimensional Ising system at $T = 0.8T_c$, 
driven by a square-wave oscillating field of amplitude $H_0 = 0.3J$. 
After \cite{KORN00}. 
({\bf a}) 
The dynamic order parameter, $\langle |Q| \rangle$, 
versus the dimensionless period $R$ for several system sizes. 
({\bf b}) 
The order-parameter fluctuation strength, $X_L^Q$, 
versus $R$ for several system sizes 
}
\label{fig:dpt}
\end{figure}
Thus, low frequencies lead to symmetric hysteresis loops, 
such as in Fig.~\ref{fig:loop}, while high frequencies produce 
loops in which the magnetization oscillates about one or the other of its 
zero-field equilibrium values. 
For small systems or weak field 
amplitudes, such that the magnetization reversal occurs via the single-droplet 
mechanism, this results in stochastic resonance \cite{SIDE98A}. 

For large systems or stronger fields, such that the magnetization switching 
occurs via the multidroplet mechanism, the transition from symmetric to 
asymmetric hysteresis loops becomes a genuine critical 
phenomenon at a sharply defined critical frequency. The transition is 
essentially due to a competition between two time scales: the 
metastable lifetime, $\tau(H_0,T)$, and the frequency of the applied field, 
$\omega/ 2 \pi$. 
As a result, the critical value of the reduced frequency $1/R$ is 
on the order of unity. This nonequilibrium phase 
transition was first observed in numerical solutions of mean-field 
equations of motion for ferromagnets in oscillating fields \cite{TOME90,MEND91}.
Subsequently it has been observed in 
numerous Monte Carlo simulations of kinetic Ising systems 
\cite{SIDE99,KORN00,LO90,ACHA94,ACHA95,ACHA97C,ACHA97D,ACHA98,CHAK99,%
BUEN98,BUEN00,SIDE98} 
and in further mean-field studies \cite{ACHA95,ACHA97D,ACHA98,BUEN98,ZIMM93A}. 
It may also have been experimentally observed in ultrathin films of Co on 
Cu(100) \cite{JIAN95,JIAN96}. 

In this far-from-equilibrium phase transition the role of order parameter is 
played by the period-averaged magnetization, 
$Q = (\omega / 2 \pi) \oint m(t) {\rm d}t$. This quantity is shown 
in Fig.~\ref{fig:dpt}({\bf a}) versus the dimensionless period $R$ for several 
system sizes. 
The order-parameter fluctuation strength, 
$X_L^Q = L^2 \left[ \langle Q^2 \rangle - \langle |Q| \rangle^2 \right]$, 
which corresponds to the susceptibility in an 
equilibrium system, is shown for several values of $L$ in 
Fig.~\ref{fig:dpt}({\bf b}). 
Both the order parameter and its fluctuations depend on $L$ in a 
way very similar to data from 
simulations of {\it equilibrium\/} phase transitions. 
And, indeed, formal 
finite-size scaling\index{finite-size scaling} analysis of 
the Monte Carlo data \cite{SIDE99,KORN00,SIDE98}, as well as  
analytical arguments \cite{GRIN85,FUJI01}, 
have shown that this far-from-equilibrium phase transition belongs to the same 
universality class as the equilibrium phase transition in the 
Ising model in zero field. This is a quite remarkable result, as it extends the 
scope of an equilibrium universality class to a far-from equilibrium system.

\section{Summary}

In this article we have presented numerical and theoretical results on
magnetization reversal and hysteresis in models of magnetic
nanoparticles and ultrathin films. Models that were explicitly
considered are kinetic Ising and classical Heisenberg models, which
were studied by dynamic Monte Carlo simulations, and continuum-spin
micromagnetics models, which were studied by finite-temperature
Langevin-equation methods. The simulation results were interpreted
within the context of nucleation theory, and it was shown how the
reversal modes change from single-droplet to multidroplet upon
increasing the strength of the applied field or the size of the system. 
Computer simulations of model
systems such as those presented here enable one to study in detail the
statistical properties of the reversal processes, as well as the
time dependent internal magnetization structure. Such simulation results
have now attained sufficient quality that they can fruitfully be compared with
present and future experiments. 

\section*{Acknowledgments}

We are happy to acknowledge the collaborators in our studies of 
magnetization switching phenomena: 
M.~Kolesik, 
G.~Korniss, 
H.L.\ Richards,
S.W.\ Sides, 
D.M.\ Townsley,
and 
C.J.\ White. 
We also thank S.~Wirth and S.~von~Moln{\'a}r for useful conversations, 
and D.D.\ Awschalom and J.~Shi for the image 
data on which Fig.~\ref{fig:awschalom} is based. 

Supported in part by U.S.\ National Science Foundation Grant No.\ DMR-9871455, 
and by Florida State University through the Center for Materials 
Research and Technology and the School of Computational Science and Information 
Technology. Supercomputer time was provided by Florida State University and 
by the U.S.\ Department of 
Energy through the National Energy Research Scientific Computing Center 
(DOE-AC03-76SF00098). 


%

\end{document}